\documentclass{article}

\begin{document}

\title{
Phase space path integral in curved space
\vskip1cm}

\author{\noindent Rafael Ferraro$^{ 1, 2,}$\thanks{ferraro@iafe.uba.ar}
\ and Mauricio Leston$^{ 1,}$\thanks{mauricio@iafe.uba.ar}
\\ \\ $^1\ $Instituto de Astronom\'\i a y F\'\i sica
del Espacio,\\ Casilla de Correo 67, Sucursal 28, 1428 Buenos
Aires, Argentina\\ \\ $^2\ $Departamento de F\'\i sica, Facultad
de Ciencias Exactas y Naturales,\\ Universidad de Buenos Aires,
Ciudad Universitaria, Pabell\' on I,\\ 1428 Buenos Aires,
Argentina}

\date{}

\maketitle

\vskip2cm

\begin{abstract}
Phase space path integral is worked out in a riemannian geometry,
by employing a prescription for the infinitesimal propagator that
takes riemannian normal coordinates and momenta on an equal
footing. The operator ordering induced by this prescription leads
to the DeWitt curvature coupling in the Schr\"{o}dinger equation.

\vskip 1cm


\vskip1cm

\end{abstract}

\vskip  1cm

\newpage

{\bf 1. Introduction}

\smallskip The research on path integral in curved space began with
DeWitt\cite{dw}, who found that the measure of
Pauli-Morete\cite{mor}, adapted for curvilinear coordinates, leads
to a coupling with the curvature in the Schr\"odinger equation
(see also Ref.\cite{cheng}). Later on Parker\cite{p} shown that
one can take profit of the ambiguous integration measure in the
configuration space to generate different values for the coupling
constant in the wave equation. This coupling can be understood as
a contribution of an ambiguous measure, which can be written in an
exponential way to dress it up as a potential $\xi\hbar^2 R$ in
the action. The ambiguity in the choice of the measure is nothing
but the reflection of the different operator orderings for the
Hamiltonian operator in curved space.

The ambiguities characterizing path integration methods in curved
space, can be reexamined in the context of phase space variables.
In phase space the integration measure seems to be better
established: the Liouville measure $dq\ dp$ is privileged because
it is invariant under canonical transformations. However, as was
demonstrated by Kucha{\v r}\cite{k}, ambiguities are still present
in phase space. Namely, the prescription for skeletonizing the
canonical functional action fails in being clean, as it is in
configuration space. So different prescriptions again lead to
different curvature coupling in the wave equation (see also
Ref.\cite{whe} , and Ref.\cite{fe} for the handling of a
relativistic system). In order to fix the ambiguities, it has been
proposed \cite{fe2} that the {\it natural} skeletonization of the
canonical functional action should be one managing canonical
coordinates and momenta on an equal footing. The skeletonized
canonical functional action proposed in Ref.\cite{fe2} is based on
those complete solutions of the Hamilton-Jacobi equation that
accomplish boundary conditions which are symmetric in coordinates
and momenta. The aim of the present paper is to apply the results
of Ref.\cite{fe2} to a phase space path integral in curved space,
and compare the result with DeWitt and Parker prescriptions for
configuration space path integration. In Section 2 we review
Parker's path integral in configuration space. In Section 3 we
present the phase space path integral proposed in Ref.\cite{fe2}.
In Section 4 we apply this later prescription to the system
studied by DeWitt and Parker. In Section 5 we display the
conclusions.

\vskip1cm

{\bf 2. Configuration space path integral}

\smallskip The propagator $K(q^{\prime}\ t^{\prime }\mid\, q\ t)$ is a two
point function that builds those solution $\Psi (q^{\prime},
t^{\prime })$ of the Schr\"{o}dinger equation satisfying a given
initial condition $\Psi_o (q)$ at time $t$:

\begin{equation}
\Psi (q^{\prime}, t^{\prime }) =\int dq\ K(q^{\prime}\ t^{\prime
}\mid\, q\ t)\ \Psi_o \left( q\right) \label{definicion}
\end{equation}

\smallskip

\noindent Thus the propagator itself must be a solution of the
Schr\"{o}dinger equation, accomplishing the boundary condition

\begin{equation}
\lim_{t^{\prime}\rightarrow t^+}K(q^{\prime}\ t^{\prime}\mid\, q\
t) = \delta(q^{\prime}, q), \label{contorno}
\end{equation}

\smallskip

\noindent and the composition rule

\begin{equation}
K(q^{\prime \prime }\ t^{\prime \prime }\mid  q^{\prime }\
t^{\prime })\ =\ \int K(q^{\prime \prime }\ t^{\prime \prime }|\,
q\ t)\ dq\ K(q\ t\ |\, q^{\prime }\ t^{\prime }). \label{comp}
\end{equation}

\smallskip

When curvilinear coordinates are used, one should care about the
geometrical character of the quantities involved in
Eq.(\ref{definicion}). In fact, the inner product

\begin{equation}
\langle\Psi_1,\Psi_2\rangle=\int dq\ \Psi_1^*(q)\
\Psi_2(q)\label{productointerno}
\end{equation}
.
\smallskip

\noindent must be invariant under general coordinate changes. So
the wave functions must be regarded as densities of weight 1/2, in
order that $dq\ \Psi_1^*(q)\ \Psi_2(q)$ results to be an invariant
volume. Some people prefer regarding the wave functions as
invariant; then they add a factor $\sqrt{g}$ in the inner product
($g$ is the determinant of the metric). The latter choice amounts
the replacement $\Psi\ \rightarrow g^{1/4}\Psi$, and the
consequent redefinition of the operators. In this paper we will
attach ourselves to the former choice. Thus, we will regard the
propagator as a bidensity of weight 1/2.

\medskip

The propagator can be formally expressed as a path integral
\cite{fh}

\begin{equation}
K(q^{\prime}\ t^{\prime}|\, q\ t)\ =\ \int {\cal D} q(t)\ \exp
\left[ {\frac i\hbar } \ S[q(t)]\right] , \label{1}
\end{equation}

\smallskip

\noindent where the integration is performed on all paths joining
the arguments of the propagator. This functional integration can
be dismantled into infinitesimal (short time) propagators
$K_\varepsilon (q^\prime, q)\equiv K(q^{\prime}\ t+\varepsilon\,
|\, q\ t)$, which can be assembled by means of the composition
rule (\ref{comp}), so rebuilding the finite time propagator. Each
infinitesimal piece must be a bidensity of weight 1/2. This {\it
skeletonization} process give sense to the otherwise ill-defined
functional integration (\ref{1}). However the choice for the
infinitesimal pieces is not unique. Thus path integration is an
ambiguous operation: each choice for the skeletonization amounts a
different quantization of the system, all of them possessing the
same classical limit.

\medskip

The infinitesimal propagator $K_\varepsilon (q^{\prime}, q)$ is
defined by developing the time evolution of the wave function in
the following way:

\begin{equation}
\Psi( q^{\prime}, t)+ \varepsilon\ \frac\partial{\partial
t}\Psi(q^{\prime},t) =\int d^n q\ K_\varepsilon (q^{\prime}, q)\
\Psi(q, t)+ O(\varepsilon^2).\label{uno}
\end{equation}

\smallskip

One can recover the wave equation in Eq.(\ref{uno}) by retaining
only the terms that are linear in $\varepsilon$. For this aim one
 replaces $\Psi(q, t)$ in the integrand of the Eq.(\ref{uno}) by
the expansion

\begin{equation}
\Psi(q, t) =
 \Psi(q^{\prime},t)
+\frac {\partial\Psi} {\partial q^{\prime\, i}} ( q^{\prime}, t)\
\xi ^i + \frac 12\frac {\partial\Psi} {\partial \ q^{\prime\,
i}\partial q^{\prime\, j}}(q^{\prime}, t)\ \xi ^i\ \xi ^j +
O(\xi^3),\label{dos}
\end{equation}

where $\xi^i=q^i-q^{\prime\, i}$.

\medskip

If the system io governed by the Lagrangian

\begin{equation}
L={1\over 2}\ g_{ij}(q)\ {\dot q}^i\ {\dot q}^j,\ \ \ \ \ \ \ \
i=1,...,n. \label{lagrangiano}
\end{equation}

\smallskip

then the Schr\"{o}dinger equation io a second order invariant
equation in curved space, whose general form is:

\begin{equation}
ih\frac \partial {\partial t}\Psi (q,t)=-\frac 12\, \hbar ^2\,
\triangle ^{(1/2)}\,\Psi (q,t)+\mu\hbar ^2\, R\, \Psi (q,t)
\label{ECUACIONGENERAL}
\end{equation}

\smallskip

\noindent where $\triangle ^{(1/2)}$ is the Laplacian for
densities of weight 1/2: $\triangle ^{(1/2)}=g^{1/4}\, \triangle\,
g^{-1/4}$, $R$ is the curvature escalar of the manifold and $\mu$
is a real number. Therefore, the terms $O(\xi^3)$ in Eq. (7)
cannot generate contributions linear in $\varepsilon$ when they
are replaced in Eq. (6). In Eq. (9), the coupling constant $\mu$
is specified by giving a prescription to path integrate the
propagator (5) or, equivalently, to order the hamiltonian
operator.

\medskip

Parker's infinitesimal propagator for the system (8) can be
expressed in the following way \cite{p}:

\[
K_\varepsilon(q^{\prime}\, \mid\, q)= \varepsilon^{n(p-1/2)}\frac
{g(q^\prime)^{1/4}g(q)^{1/4}}{\left( 2\pi i\hbar \right)
^{n/2}}\left( g(q^{\prime}) ^{-1/2}\ D_\varepsilon(q^{\prime},
\mid q)\ g(q)^{-1/2}\right)^p\]
\begin{equation}
\ \ \ \ \ \ \ \ \ \ \ \ \ \ \ \ \ \exp \left[ \frac i\hbar
S_\varepsilon(q^{\prime},\mid q)\right]\label{parker}
\end{equation}

\smallskip

\noindent where $D(q^{\prime}t^{\prime}\mid q\ t)$ is the Van
Vleck determinant \cite{vv}

\begin{equation}
D(q^{\prime}t^{\prime}\mid q\ t) \equiv \det \left( -\frac{
\partial ^2 S(q^{\prime}t^{\prime },q
t) }{\partial q^{\prime}\ \partial q}\right),\label{vanvleck}
\end{equation}

\smallskip

\noindent and $S(q^{\prime }\ t^{\prime }\mid q\ t)$ is the
Hamilton principal function, i.e. the functional action $S=\int L\
dt$ evaluated on its stationary (classical) path.

\medskip

Parker's propagator (\ref{parker}) satisfies the wave equation

\begin{equation}
ih\frac \partial {\partial t^{\prime}}K(q^{\prime}\
t^{\prime}\mid\, q\ t)=\Big[-\frac 12\hbar ^2\Delta^{\prime\
(1/2)}\ +\ {1\over 6}\ (1-p)\ \hbar ^2R(q^{\prime})\Big]\
K(q^{\prime}\ t^{\prime}\mid\, q\ t).\label{ECUACIONGENERAL}
\end{equation}

\smallskip

The choice $p=1/2$ in Eq. (\ref{parker}) is the DeWitt's
propagator \cite{dw}, while $p=0$ corresponds to the Feynman's
propagator \cite{f}.

\medskip

By composing the infinitesimal pieces (\ref{parker}) one gets a
meaning for the path integral (\ref{1}). Thus the Van Vleck
determinants take part in the measure $\mathcal{D}\mathbf{q}(t)$
and the Hamilton principal functions $S \left(
q^{\prime}t^{\prime}\mid q t\right)$ are the pieces that
skeletonize the functional action $S[q(t)]$:

\begin{equation}
S[q(t)]\ \longrightarrow\ \sum_{k=0}^{N-1}S(q_{k+1}\ t_{k+1}\mid
q_k\ t_k), \label{skc}
\end{equation}

\smallskip

\noindent where $t_{k+1}-t_k=\varepsilon$, and $q_k$ is a
shorthand for $q^i(t_k)$.

\vskip1cm

{\bf 3. Phase space path integral}

\smallskip Phase space path integral,

\begin{equation}
K(q^{\prime}\ t^{\prime}|\, q\ t)\ =\ \int {\cal D}p(t)\ {\cal D}
q(t)\ \exp \left[ {\frac i\hbar }\ S[q(t), p(t)]\right] ,
\label{(4)}
\end{equation}

\smallskip

\noindent also requires a skeletonized version to make sense. The
infinitesimal propagator proposed in Ref.\cite{fe2} is
\[
K_\varepsilon (q^{\prime}, q)=K(q^{\prime \prime }\ t^{\prime
\prime }=t^{\prime }+\varepsilon \;|\; q^{\prime }\ t^{\prime
})=\int {\frac{d^np_a}{(2\pi \hbar )^n}}\ \left| \ \!{}\det
\frac{\partial ^2J(q^{\prime \prime }\,t^{\prime \prime }\,|\, p\,
t)}{\partial \,q^{\prime \prime \,\,j}\;\partial \,p_a\;}\ \right|
^{\ 1/2\ }\]
\begin{equation}
 \left| \ \!{}\det -{\frac{\partial ^2J(p\, t\,|\, q^{\prime }\,t^{\prime
})}{\partial \,p_a\;\partial \,q^{\prime \,\,j}\;}}\ \right| ^{\
1/2} \nonumber \\    \nonumber \\ \exp \left[ {\frac i\hbar }\
\left( J(q^{\prime \prime }\,t^{\prime \prime }\,|\, p\, t)+J(p\,
t\,|q^{\prime }\,t^{\prime })\right) \right] \ , \label{9}
\end{equation}

\medskip

\noindent Here $J(q\,t^{\prime}\,|\, p\, t)$ and $J(p\,
t\,|q\,t^{\prime })$ are two invariant functions related to the
dynamical system, which should take on an equal footing canonical
coordinates and momenta. Besides these functions should provide a
skeletonized version of the canonical functional action $S[q(t),
p(t)]$:

\begin{equation}
S[q(t), p(t)]\longrightarrow \sum_{k=0}^{(N-2)/2}\left\{
J(q_{2k+2} t_{2k+2}|p_{2k+1}t_{2k+1})+J(p_{2k+1} t_{2k+1}|q_{2k}
t_{2k})\right\} .  \label{3'}
\end{equation}

\smallskip

The functions $J(q\,t^{\prime}\,|\, p\, t)$ and $J(p\,
t\,|q\,t^{\prime })$ used in Ref.\cite{fe2} are two complete
solutions of the Hamilton-Jacobi equation linked by the relation

\begin{equation}
J(q\,t^{\prime}\,|\, p\, t)=-J(p\, t\,|q\,t^{\prime }),
\label{relacion}
\end{equation}

\smallskip

The invariance of $J(q\,t^{\prime}\,|\, p\, t)$ is not guaranteed
by the invariant Hamiltonian in the Hamilton-Jacobi equation,
because the boundary condition must also be invariant. The
boundary condition proposed in Ref.\cite{fe2} is

\begin{equation}
J(q^i\, t\,|p_a\, t)=p_a\ \sigma^a(q^j),  \label{181}
\end{equation}

\smallskip

\noindent $\sigma^a(q^j)$ in Eq.(\ref{181}) are riemannian normal
coordinates for the point $P$ identified by the coordinates $q^i$,
based at some origin $O$:

\begin{equation}
\sigma^a(q^j)= s(O,P)\ u^a,  \label{182}
\end{equation}

\smallskip

\noindent where $s(O,P)=\int_O^P \sqrt{ g_{ij}dq^idq^j}$ is the
(invariant) lenght of the geodesic joining $O$ and $P$, and $u^a$
are components of the unitary vector at $O$ that is tangent to the
geodesic joining $O$ and $P$, in some basis chosen at $O$.
Riemannian normal coordinates have a twofold behavior: like $s$
they are invariant under changes of coordinates $q^i$, but
transform as components of a vector under changes of basis in the
target space $T_O$. $p_a$ in Eqs.(\ref{9}) and (\ref{181}) are
momenta canonically conjugated to $\sigma^a(q^j)$,

\begin{equation}
p_j=\frac{\partial \sigma ^a}{
\partial \,q^j}(q^i)\;p_a\label{183}
\end{equation}

\smallskip

\noindent Thus the boundary condition (\ref{181}) is invariant
under both general changes of the $q^i$'s and change of basis in
the target space $T_O$. The boundary condition (\ref{181}) means
that canonically conjugated riemannian normal coordinates and
momenta enter the skeletonization on an equal footing.

\smallskip

As is shown in Ref.\cite{fe2}, $J(q^{\prime}\,t^{\prime}\,|\, p\,
t)$  and $J(p^{\prime }\, t^{\prime }\,|q\,t)$, there named Jacobi
principal functions, result to be the Legendre transforms of
$S(\sigma^a(q^{\prime})\,t^{\prime}\,|\, \sigma^a(q)\, t)$ that
interchange the variables $\sigma^a$ and $p_a$ at $t$ and
$t^{\prime}$ respectively. Then $J(q^{\prime}\,t^{\prime}\,|\, p\,
t)$ generates contact transformations

\begin{equation}
p_j^{\prime  }\ =\ {\frac{\partial J(q^{\prime\,i}\ t^{\prime
}|p_a\ t)}{\partial \,q^{\prime \,j}\ }} \ \ ,\ \ \ \ \ \ \ \ \ \
\ \ \ \ \ \ \sigma^a\ =\ {\ \frac{\partial J( q^{\prime\,i}\
t^{\prime }|p_a\ t)}{\partial \,p_a}}. \label{(5')}
\end{equation}

\smallskip

\noindent So the determinants in the infinitesimal propagator
(\ref{9}) are well defined in all phase space. The functions
$J(q\,t^{\prime}\,|\, p\, t)$ and $J(p\, t^{\prime }\,|q\,t)$ also
provides a proper skeletonization for the canonical action:

\[
J(q_{2k+2}\ t_{2k+2}\ |p_{2k+1}\ t_{2k+1})+J(p_{2k+1}\ t_{2k+1}|
q_{2k}\ t_{2k})  \]
\begin{equation}
\longrightarrow \  p_{a\ 2k+1}\ (\sigma_{2k+2}^a - \sigma_{2k}^a)
- H(\sigma^a_{2k}, p_{j\ 2k+1}, t_{2k+1})\,(t_{2k+2} - t_{2k}).
\label{(91)}
\end{equation}

\smallskip

Differing from the infinitesimal propagator (\ref{parker}), which
only holds for systems whose Lagrangian is quadratic in the
velocities \cite{sch}, the infinitesimal propagator (\ref{9}) can
be applied to any hamiltonian system. It will provide a unitary
evolution when $t$ in Eq. (\ref{9}) is the mid time $t\equiv
t^{\prime }+(\varepsilon \,/2)=t^{\prime \prime }-(\varepsilon
\,/2)$.

Finally we remark that it is by no means evident that the
prescription (\ref{9}) does not depend on the choice of the origin
$O$ for defining normal riemannian coordinates. This topic will be
analyzed in the next Section.

\vskip1cm

{\bf 4. Infinitesimal propagator in the riemannian geometry}

\smallskip

Let be $g_{ab}(q)$ the components of the metric tensor at $P$ in
the riemannian normal coordinates system (RNCS),

\begin{equation}
g_{ab}(q)\equiv \frac{\partial q^j}{\partial \sigma ^a}(q)\
\frac{\partial q^i}{\partial \sigma ^b}(q)\ g_{ij}(q)\qquad
g^{ab}(q)\equiv \frac{\partial \sigma ^a} {
\partial q^i}(q)\ \frac{\partial \sigma ^b}{\partial q^j}(q)\ g^{ij}(q)
\end{equation}

\smallskip

\noindent $g_{ab}$ is invariant under changes of $q^i$'s, but is a
tensor in $T_O$. Since $u^a$ in Eq.(\ref{182}) is unitary
($g_{ab}u^a u^b=1$), one recognizes that

\begin{equation}
g_{ab}\ \sigma^a\ \sigma^b = s^2.
\end{equation}

\smallskip

Then,

\begin{equation}
\sigma^a (q) = \frac 12 g^{ab}(q) \frac{\partial
s^2}{\partial\sigma^b}(q)= s\ g^{ab}(q) \frac{\partial
s}{\partial\sigma^b}(q),
\end{equation}

\smallskip

and

\begin{equation}
u^a (q) = g^{ab}(q)\ \frac{\partial s}{\partial\sigma^b}(q).
\end{equation}

\medskip

At the lowest order in $\varepsilon=\Delta t$, the solution of the
Hamilton-Jacobi equation fulfilling the boundary condition
(\ref{181}) is

\begin{equation}
J_\varepsilon(q^i\, \mid p_a) = p_a\sigma^b(q^i)-\frac 12\ g^{ab}\
p_a\ p_b\ \varepsilon + O(\varepsilon^2),\label{INFINITESIMAL}
\end{equation}

\medskip

\begin{equation}
\left| \frac{\partial^2 J_{(\varepsilon/2)}(q^i,
p_a,\varepsilon)}{\partial q^i\partial p_a}\right| ^{1/2}=\left|
\frac{\partial \sigma ^a}{\partial q^i}\right| ^{1/2}\left(
1-\frac 14\frac{
\partial g^{ab}}{\partial \sigma^a}\ p_b\ \varepsilon\right)+
O(\varepsilon^2)\label{jac}
\end{equation}

\bigskip

\noindent Both Jacobi principal functions, the one of
Eq.(\ref{INFINITESIMAL}) and its companion of Eq.(\ref{relacion}),
enter the Eq.(\ref{9}) to compute the infinitesimal propagator.
The result is (see Appendix)
\[
K_\varepsilon \left( q^{\prime \prime }\, t^{\prime \prime
}=t^\prime + \varepsilon\, |\, q^{\prime }\, t^{\prime }\right) =
\frac {\varepsilon^{n/2}}{\left( 2\pi i\hbar \right) ^{n/2}}\
[\det g_{ij}( q^{\prime \prime })]^{-1/4}\ D_\varepsilon(
q^{\prime \prime }\, |\, q^{\prime })\ [\det g_{ij}( q^\prime)
]^{-1/4}\]

\begin{equation}
\exp \frac i\hbar \left[ S_\epsilon\left( q^{\prime \prime }\, |\,
q^{\prime }\right) -\frac{\hbar ^2}{2}\int\nolimits_{t^{\prime
}}^{t^{\prime \prime }}dt\ (\det g_{ab})^{1/4}\triangle\left[
(\det g_{ab})^{-1/4}\right]\right],\label{PROPAGADORNUESTRO}
\end{equation}

where the integral on $t$ is evaluated along the classical
trajectory $q(t)$ joining $(q^\prime, t^\prime)$ and
$(q^{\prime\prime}, t^{\prime\prime})$. This infinitesimal
propagator leads to the following wave equation:

\begin{equation}
ih\frac \partial {\partial t}\Psi (q,t)=-\frac{\hbar ^2}{2}
\triangle ^{(1/2)}\,\Psi (q,t)+\frac 12\hbar ^2(\det
g_{ab})^{\frac 14}\triangle  \left[(\det(g_{ab}))^{-\frac
14}\right]\, \Psi (q,t). \label{ECUACION}
\end{equation}

The term $M_O(q) \equiv $ $\frac{\hbar ^2}{2} (\det
g_{ab}(q))^{\frac 14}\triangle  \left[ (\det g_{ab}(q))^{-\frac
14}\right]$ is  invariant under changes of the coordinates of
point $P$, where the laplacian operates, and also under change of
basis in $T_O$. Except for the factor $\exp \left[-\frac
i\hbar\int\nolimits_{t^{\prime }}^{t^{\prime \prime }}dt\
M_O\left(q(t)\right)\right]$, the result (\ref{PROPAGADORNUESTRO})
agrees with (\ref{parker}) for $p=1$. However this factor depends
on the choice of the origin $O$, which is a non admissible
dependence for the propagator. To be convinced that $M_O$ really
depends on $O$, one can evaluate $M_O$ in RNCS:

\[
(\det g_{ab})^{1/4}\triangle \ (\det g_{ab})^{-1/4}=
\]
\[
(\det g_{ab})^{-1/4}\frac
\partial {\partial \sigma ^a}(\det g_{ab})^{1/2}g^{ab}\frac{\partial
(\det g_{ab})^{-1/4}}{\partial \sigma ^b}=
\]
\begin{equation}
-\frac 14g^{ab}g^{cd}\frac{\partial ^2g_{cd}}{\partial \sigma
^a\partial\sigma ^b} -\frac 1{16}g^{cd}\frac{\partial g_{cd}}{\partial \sigma ^a}%
g^{ab}g^{ef}\frac{\partial g_{ef}}{\partial \sigma ^b}-\frac 14\frac{%
\partial g^{ab}}{\partial \sigma ^a}g^{cd}\frac{\partial g_{cd}}{\partial
\sigma ^b}
\end{equation}

This expression can be compared with the curvature scalar $R(q)$:

\[
R=\frac 12\frac{\partial g_{ab}}{\partial\sigma^c}\ \frac{\partial
g_{de}}{\partial\sigma^f}\ \left(
g^{ac}g^{de}g^{bf}-g^{ab}g^{de}g^{cf}+2g^{dc}g^{be}g^{af}+2g^{df}g^{be}g^{ac}\right)
\]

\begin{equation}
+\frac{\partial^2 g_{ab}}{\partial\sigma^c\partial\sigma^d}\left(
g^{ac}g^{bd}-g^{ab}g^{cd}\right) \label{CURVATURA}
\end{equation}

Since they do not coincide, the term $M_O$  depends on the choice
of the origin $O$ to define the $p_a$'s ( in flat space $M_o$ is
null because RNCS is the cartesian system where the derivatives of
the metric are null) .

\vskip1cm

\noindent{\bf 5. Prescription for the origin of RNCS in the
infinitesimal propagator}

\smallskip

In order to avoid any undesirable dependence on geometric objects
non linked to the physical system, the point $O$ in
(\ref{PROPAGADORNUESTRO}) should be dictated by the dynamical
system itself. So the point $O$ for the infinitesimal propagator
(29) must be chosen on the classical path joining $q^{\prime
\prime }(t^{\prime \prime })$ and $q^{\prime }(t^{\prime })$. This
prescription will give a proper meaning to the factor $\exp
\left[-\frac i\hbar\int\nolimits_{t^{\prime }}^{t^{\prime \prime
}}dt\ M_O\left(q(t)\right)\right]$ in Eq.
(\ref{PROPAGADORNUESTRO}). The simplest prescription is to choose
the point $O$ as the final point in the infinitesimal propagator.
So $M_O(q)$ in Eq. (\ref{ECUACION}) must be replaced by $M_P(q)$,
being $P$ the point where the wave function is evaluated. Having
into account the expansion (\cite{k})

\begin{equation}
 \det (\frac {\partial\sigma^a}{\partial q^j}) =g^{-\frac 12}(P)
 g^{-\frac 12}(O)\left( 1+\frac 16 s^2R^{ij}s
 _{,i} s_{,j}+O\left( s ^6\right) \right),
\end{equation}

\smallskip

\noindent then

\begin{equation}
\lim_{O\rightarrow P} M_O(q) =\frac {\hbar ^2}{12}\ R(q).
\label{LIMITE}
\end{equation}

\vskip1cm

\noindent Thus the Schr\"{o}dinger equation(30)turns out to be:

\begin{equation}
ih\frac \partial {\partial t}\Psi (q,t)=-\frac 12\hbar ^2\triangle
^{(1/2)}\,\Psi (q,t)+\frac 1{12}\hbar ^2R\Psi (q,t)
\label{ECUACIONPOSTA}
\end{equation}

\noindent which exhibits the curvature coupling prescribed by
DeWitt.

\vskip1cm

\noindent{\bf 6. Conclusions }

\smallskip

To some extent the prescription $O\equiv P$ in the infinitesimal
propagator seems arbitrary, in the sense that any other point of
the classical trajectory joining $q^{\prime}$ and
$q^{\prime\prime}$ is equally natural to play that role. However,
the result of Section 5 -the DeWitt curvature coupling- is not
affected by a displacement of point $O$ along the classical path.
In fact, such a displacement would change $M_o(q(t))$ to
$M_o(q(t))+ O(\xi^a)$; thus
$K_\varepsilon\left(q^{\prime\prime}\mid q^{\prime}\right)$ would
get a contribution that can be expanded in powers of $\xi^a$.
Because of the rule $\xi^a\ \xi^b\longrightarrow i\, \varepsilon\
\hbar\ g^{ab}$ (see Eq. (\ref{V}) in the Appendix), these changes
do not modify the leading term in Eq. (6) but only contribute to
higher order in $\varepsilon$. Then, these changes can be ignored
at the level of the infinitesimal propagator. Therefore the DeWitt
curvature coupling appears to be the natural one for a quadratic
system.

\vskip2cm

\noindent{\bf Appendix }

\vskip1cm

{\bf I Identities }

\medskip

\begin{equation}
g_{ij,k}=-g_{ih}\ g^{hl},_kg_{lj}  \label{I}
\end{equation}

\begin{equation}
g^N,_i=Ng^Ng^{mn}g_{nm,i}  \label{II}
\end{equation}

\begin{equation}
\int\nolimits_{-\infty }^\infty\ \frac{d^nx}{\left( 2\pi \right)
^{n/2}}\exp \left[ -\frac 12\left( x,Ax\right) +\left( b,x\right)
\right] =\left( \det A\right) ^{-1/2}\exp \frac 12\left(
b,A^{-1}b\right) \label{III}
\end{equation}

\medskip

\noindent where $A$ is a $n\times n$ matrix and $b$ and $x$ are
vectors.

\bigskip

\[
\int\nolimits_{-\infty }^\infty \triangle q^{\alpha _1}\triangle
q^{\alpha _2}...\triangle q^{\alpha _{2m}}\exp \left( \frac
i{2\hbar \varepsilon }g_{ij}\triangle q_i\triangle q_j\right)
d^nq=\left( 2\pi \hbar \varepsilon \right) ^{N/2}g_{}^{-1/2}\left(
i\hbar \varepsilon \right) ^m
\]

\begin{equation}
\{g^{\alpha _1\alpha _2}...g^{\alpha _{2m-1}\alpha _{2m}}\}+ \ \
all\ \ relevant\ \ permutations\ \ of\ \  \alpha _1,...,\alpha
_{2m}  \label{V}
\end{equation}

\vskip1cm

{\bf  II Derivation of Wave Equation }

\medskip

The expression (\ref{9}) can be integrated with the Jacobi
principal function (27). The result is:

\[
K_\varepsilon\left( q^{\prime \prime }\mid q^{\prime } \right)
=\frac 1{\left( 2\pi\hbar\varepsilon \right) ^{n/2}}g^{\prime
\frac 14 }g^{\prime\prime \frac 14}(\det(g_{ab}^{\prime \prime}
))^{-\frac 14}(\det(g_{ab}^{\prime } ))^{-\frac
14}(\det(\overline{g}_{ab}))^{1/2}\]
\begin{equation}
\left[ 1+\frac 14\overline{g}_{ab}\left( \frac{%
\partial g^{\prime bd}}{\partial \sigma ^{\prime d}}-\frac{\partial
g^{\prime \prime bd}}{\partial \sigma ^{\prime \prime b}}\right)
\xi ^a\right]\ \exp \left(\frac i{2\hbar \varepsilon
}\overline{g}_{ab}\xi ^a\xi ^b\right)
\end{equation}

\bigskip

\noindent where $\overline{g}^{ab}\equiv \frac 12\left( g^{\prime
\prime ab}+g^{\prime ab}\right) $ .  Let be $\xi ^j\equiv \sigma
^{\prime \prime \,j}-\sigma ^{\prime \,j}$ . Then:

\begin{equation}
\overline{g}_{ab}\mid _{\xi =0}=g_{ab}
\end{equation}

\begin{equation}
\frac{\partial \overline{g}_{ab}}{\partial \sigma ^{\prime c}}\mid
_{\xi =0}=\frac 12\frac{\partial g_{ab}^{\prime  }}{\partial
\sigma ^{\prime c}}
\end{equation}

\begin{equation}
\frac{\partial \overline{g}^{ab}}{\partial \sigma ^{\prime c}}\mid
_{\xi =0}=\frac 12\frac{\partial g^{\prime  ab}}{\partial \sigma
^{\prime  c}}
\end{equation}

\begin{equation}
\frac{\partial ^2\overline{g}_{ab}}{\partial \sigma ^{\prime
c}\partial \sigma ^{\prime d}}\mid _{\xi=0}=\frac 12\frac{\partial
^2g_{ab}^{\prime \prime }}{\partial \sigma ^{\prime \prime
c}\partial \sigma ^{\prime \prime d}}+\frac 14\left(
\frac{\partial g_{am}^{\prime \prime }}{\partial \sigma ^{\prime
\prime c}}\frac{\partial g^{\prime \prime mn}}{\partial \sigma
^{\prime \prime d}}g_{nb}^{\prime \prime }+g_{am}\frac{\partial
g^{\prime \prime mn}}{\partial \sigma ^{\prime \prime
d}}\frac{\partial g_{nb}^{\prime \prime }}{\partial \sigma
^{\prime \prime c}}\right)
\end{equation}

\bigskip

By expanding $det(g^{\prime}_{ab})$ and $det(\overline{g}_{ab})$
around $\sigma ^{\prime \prime \,a}$,

\begin{equation}
(\det g_{ab}^{ \prime} )^{-\frac 14}=(\det g_{ab}^{\prime \prime}
)^{-\frac 14}- \frac{\partial (\det(g_{ab}^{\prime \prime}
))^{-\frac 14}}{\partial \sigma ^{\prime \prime c}}\xi ^i+\frac
12\frac{\partial ^2(\det(g_{ab}^{\prime \prime} ))^{-\frac
14}}{\partial \sigma ^{\prime \prime c}\partial \sigma ^{\prime
\prime d}}\xi ^c\xi ^d + O(\xi^3)
\end{equation}

\bigskip

\noindent and using the identities (42-44), one obtains the terms
that are relevant in the integration of Eq. (6). According to
Eq.(\ref{V}) these terms are:

\begin{equation}
(\det \overline{g}_{ab})^{\frac 12}=(\det g_{ab}^{\prime \prime}
)^{\frac 14}(\det g_{ab}^{\prime } )^{\frac 14}+(\det
g_{ab}^{\prime \prime} )^{\frac 12}\frac 1{16}\frac{\partial
g^{\prime \prime am}}{\partial \sigma ^{\prime \prime d}}
\frac{\partial g_{am}^{\prime \prime }}{\partial \sigma ^{\prime
\prime c}} \xi ^c\xi ^d + O(\xi^3)
\end{equation}

\bigskip

\[
\overline{g}_{ab}\ \xi ^a\left( \frac{\partial g^{\prime
bc}}{\partial \sigma ^{\prime c}}-\frac{\partial g^{\prime \prime
bc}}{\partial \sigma ^{\prime c} }\right) =\frac{\partial
^2g_{an}^{\prime \prime }}{\partial \sigma ^{\prime \prime
c}\partial \sigma ^{\prime \prime d}}g^{\prime \prime nc}\xi
^{a}\xi ^{d}+ \]
\begin{equation}
g_{ab}^{\prime \prime }\frac{\partial g^{\prime \prime bm}}{
\partial \sigma ^{\prime \prime d}}\frac{\partial g_{mn}^{\prime \prime }}{
\partial \sigma ^{\prime \prime c}}g^{\prime \prime nc}\xi ^a\xi
^d+\frac{\partial g^{\prime \prime nc}}{\partial \sigma ^{\prime
\prime d}} \frac{\partial g_{an}^{\prime \prime }}{\partial \sigma
^{\prime \prime c}} \xi ^a\xi ^d+O\left( \xi ^3\right)
\end{equation}

\bigskip

\[
\exp \frac i{2\hbar \varepsilon }\overline{g}_{ab}\ \xi ^a\xi ^b
=\exp \left[\frac i{2\hbar \varepsilon }g_{ab}\ \xi ^a\xi
^b\right]\ \Big\{1-\frac i{4\hbar \varepsilon }\frac{
\partial g_{ab}^{\prime \prime }}{\partial \sigma ^{\prime \prime
c}}\ \xi^a\ \xi^b\ \xi^c + \]\ \[ \frac i{8\hbar \varepsilon
}\frac{\partial ^2g_{ab}^{\prime \prime }}{
\partial \sigma ^{\prime \prime c}\partial \sigma ^{\prime \prime d}}\ \xi
^a\xi ^b\xi ^c\xi ^d+\frac i{8\hbar \varepsilon }\frac{\partial
g_{am}^{\prime \prime }}{\partial \sigma ^{\prime \prime
c}}\frac{\partial g^{\prime \prime mn}}{\partial \sigma ^{\prime
\prime d}}g_{nb}^{\prime \prime }\ \xi ^a\xi ^b\xi ^c\xi ^d
\]
\begin{equation}
 -\frac 1{32\hbar ^2\varepsilon ^2}\frac{\partial g_{ab}^{\prime \prime }}{
\partial \sigma ^{\prime \prime c}}\frac{\partial g_{de}^{\prime \prime }}{
\partial \sigma ^{\prime \prime f}}\ \xi ^a\xi ^b\xi ^c\xi ^d\xi ^e\xi
^f+... \Big\}
\end{equation}

\bigskip

These expressions are replaced in Eq. (40) to obtain the
infinitesimal propagator $K_\varepsilon\left(q^{\prime\prime}\mid
q^{\prime}\right)$ entering the integration in Eq.(6). This
integration is performed with the help of the identities
(\ref{V}). Thus the wave equation results to be

\begin{equation}
ih\frac{\partial \Psi \left( q^{\prime \prime },t\right) }{
\partial t}=-\frac 12\hbar ^2\triangle ^{(1/2)}\Psi\left( q^{\prime
\prime },t\right) +\frac 12\hbar ^2\left(
(\det(g_{ab}^{\prime \prime} ))^{\frac 14}\triangle (\det
(g_{ab}^{\prime \prime} ))^{-\frac 14}\right) \Psi \left(
q^{\prime \prime },t\right)
\end{equation}

\vskip1cm

\noindent{\bf ACKNOWLEDGMENTS}

This work was supported by Universidad de Buenos Aires (Proy.
TX64) and Consejo Nacional de Investigaciones Cient\'\i ficas y
T\'ecnicas (Argentina).

\end{document}